# Multistep pulse compressor based on single-pass single-grating-pair main compressor

**SHUMAN DU**[1, 2, 3], **XIONG SHEN**[1, 3], **WENHAI LIANG**[1, 2], **PENG WANG**[1], **JUN LIU**[1, 2, *], **RUXIN LI**[1, 2, 4]

[1] *State Key Laboratory of High Field Laser Physics and CAS Center for Excellence in Ultra-intense Laser Science, Shanghai Institute of Optics and Fine Mechanics, Chinese Academy of Sciences, Shanghai 201800, China*

[2] *University Center of Materials Science and Optoelectronics Engineering, University of Chinese Academy of Sciences, Beijing 100049, China*

[3] *These authors contributed equally to this work.*

[4] *ruxinli@siom.ac.cn*

*Corresponding author: jliu@siom.ac.cn*

**Abstract:** A multistage smoothing multistep pulse compressor (MPC) based on a single-pass single-grating-pair (SSGP) main compressor is proposed to simplify the entire petawatt (PW) compressor. Only one grating pair with relatively long distance is used to generate the same amount of spectral dispersion in the main compressor compared with a four-grating main compressor. As the SSGP induces the largest spatial dispersion, it can introduce the best beam-smoothing effect to the laser beam on the last grating. When considering the diffraction loss of only two gratings, the total compression efficiency of the SSGP main compressor is even larger than that of a four-grating main compressor. Furthermore, the spatiotemporal aberration induced by single-grating-pair can be compensated effectively by using deformable mirrors, however it is difficult or complicated to be well compensated in a four-grating compressor. Approximately 50-100 PW laser pulses can be obtained using this SSGP-based multistage smoothing MPC with a single laser beam.

1. Introduction

Since the invention of lasers in 1960 [1], laser peak power has been increased continuously in two ways: reducing the pulse duration through methods such as Q-switching and mode-locking and increasing pulse energy through laser amplification. However, this increase in peak power had been stagnated for about twenty years by 1985 due to the laser crystal damage problem during the amplification processes. With the invention of the chirped pulse amplification (CPA) method [2] and the optical parametric chirped pulse amplification (OPCPA) method [3], the laser peak power continuously increased again after 1985. Currently, the two highest 10 PW lasers are in operation at SULF (China) and ELI-NP (Europe) [4-7]. According to a recent review nearly fifty petawatt (PW) level laser facilities exist worldwide [8]. Moreover, several institutions in Europe, the United States of America (USA), Russia, Japan and China have reported ambitious plans to achieve 10s-100s PW lasers based on OPCPA: ELI-200PW (Europe), EP-OPAL-75PW (USA), XCELS-200PW (Russia), GEKKO-EXA-50PW (Japan), and SEL-100PW (China). The focused intensities of these lasers are expected to be higher than $10^{23}$ W/cm$^2$ [8]. This type of ultrahigh peak power laser can push the fundamental light-electron interaction to quantum electron dynamics; trigger the creation of particles such as electrons, muons, pions and their corresponding antiparticles; facilitate nuclear quantum optics; and potentially lead to the discovery of new particles beyond the standard model [9].

For all the planned 10s-100s PW laser facilities, the main problem is the absence of compression gratings with a sufficiently high damage threshold and large enough size, where the last grating with the shortest pulse duration is the short-board of the grating-based compressor. As a result, phase-locking of multiple above 10-PW laser channels is the earliest solution to achieve 100s PW for XCELS-200PW, ELI-200PW, and SEL-100PW [8]. However, this phase-locking is very sensitive to many laser parameters across the laser channels, such as optical delay, pointing stability, beam wavefront, and spectral dispersion [10]. A method called multistep pulse compressor (MPC) has been proposed recently to solve the compression problem. In this method, the compression pulse energy problem is transferred to the spatiotemporal properties of the input/output laser beam [11]. As an improvement of the typical MPC, a multistage smoothing MPC (MS-MPC) based on an asymmetric four-grating compressor (AFGC) was also proposed to achieve ultrahigh peak power output and safe operation, simultaneously [12, 13].

In this study, the MS-MPC method was further improved based on a single-pass single-grating-pair (SSGP) main compressor. In comparison to the AFGC-based main compressor, only two parallel gratings with relatively long distance were used to generate the same amount of spectral dispersion in the main compressor. Furthermore, the grating pair can induce the largest spatial dispersion compared to the AFGC, which can achieve the best beam smoothing effect. Besides saving for two expensive gratings, the total compression efficiency of the SSGP main compressor was even larger than that of the AFGC because there are only two gratings induced diffraction loss. Moreover, the spatiotemporal aberration induced by the wavefront distortion of the large gratings in the main compressor, which will significantly reduce the final focused intensity, cannot be compensated effectively in a four-grating main compressor[14-16]. However, in the proposed MS-MPC based on an SSGP design, the wavefront distortion can be compensated directly by using deformable mirrors.

## 2. The SSGP-based main compressor

### 2.1 Optical schemes of the SSGP main compressor

In a previous MS-MPC [13], the AFGC was used as the main compressor to compensate for the spectral dispersion and simultaneously induce a suitably small spatial dispersion to smooth the laser beam. This is because the AFGC is equal to a symmetric four-grating compressor together with an SSGP compressor with a short distance, where only the SSGP compressor can induce a larger spatial dispersion. The distance may be increased between the two grating pairs, L2–L1 in Fig. 1(a), to introduce the maximum spatial dispersion width, limited by the optical block between the grating pair. The optical schemes of the AFGC and SSGP are shown in Fig. 1(a) and (b), respectively. The two compressors can induce the same amount of spectral dispersion with different amounts of spatial dispersion. The induced spatial dispersion ratio between the AFGC and SSGP is (L2-L1) /(L2+L1).

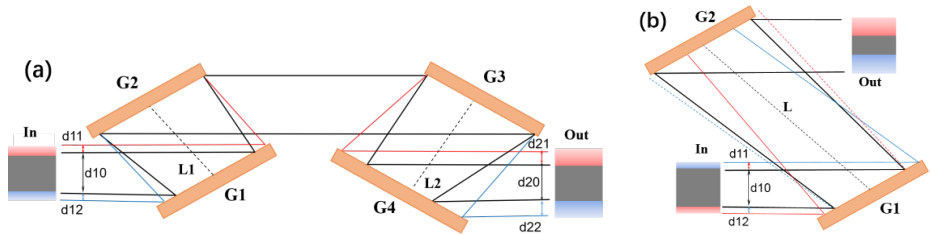

Fig.1. Optical scheme of the (a) AFGC main compressor and (b) SSGP main compressor. G1-G4 are diffraction gratings; L=L1+L2.

In a grating-pair-based pulse compressor [17], the induced phase shift can be expressed as $\Phi(\omega)= \omega D_0 [1+\cos(\theta_r-\theta_i)]\sec\theta_r/c$, where $\omega$ is the laser frequency, $D_0$ is the perpendicular distance of the grating pair, and $\theta_r$ and $\theta_i$ are the diffraction angle and the incident angle, respectively. The induced spectral dispersions are the Taylor coefficients of $\Phi(\omega)$, all linearly related to $D_0$. For example, the second-order spectral dispersion can be expressed as $\Phi_2= -8\pi^2 c D_g/(\omega_0^3 d^2 \cos^2\theta_{r0})$, where $D_g=D_0\sec\theta_{r0}$ is the central distance of the two parallel gratings, $d$ is the grating constant, and $\omega_0$ and $\theta_{r0}$ are the laser frequency and diffraction angle at the central wavelength, respectively. According to the expression, the induced spectral dispersion of an SSGP compressor can equal that of a four-grating compressor when the total grating pair central distance is equal, that is, L=L1+L2. As a result, the typical four-grating compressor can be replaced using a compressor with only two parallel gratings separated by relatively long distance. As the distance between G1 and G2 is doubled, the SSGP main compressor can be used for the dispersive compensation of even larger laser beams, which may induce a light block on the edge of G1 or G2 for a typical four-grating main compressor.

**2.2 Beam-smoothing effect of the SSGP main compressor**

Although the induced spectral dispersion is the same for the four-grating compressor and the SSGP compressor if L=L1+L2, their output laser beams have different spatial dispersions. In a typical symmetric four-grating compressor, the second grating pair compensates for the spatial dispersion induced by the first grating pair, and the induced total spatial dispersion is zero. The AFGC can induce a suitably small spatial dispersion to smooth the output laser beam in the MS-MPC. As for the SSGP compressor, it induces the maximum spatial dispersion to the output laser beam. According to previous work, the spatial dispersion induced beam smoothing effect is positively related to the induced spatial dispersion width [11-13]. From simulation results, an induced 50 mm spatial dispersion width reduces the spatial intensity modulation from 2.0 to about 1.1. The SSGP main compressor induces a spatial dispersion width of hundreds of millimeters to the output laser beam. As the induced spatial dispersion width is so large, the output laser beam is well smoothed even for a laser beam with a larger spatial intensity modulation or wider spatial modulation frequency.

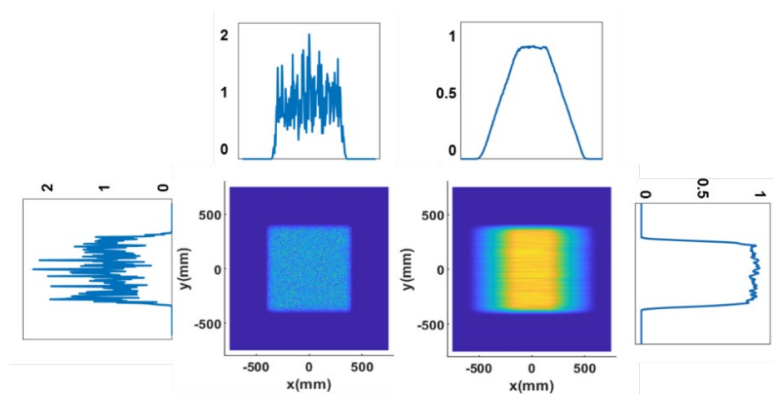

Fig.2. Spatial profiles of the modulated input and the smoothed output laser beams.

A pulse centered at 925 nm with a 200 nm full spectral bandwidth, 10$^{th}$ order super-Gaussian profile, and 860×860 mm$^2$ beam size, was used as the simulation input laser source to check the beam smooth effect of the SSGP compressor. Figure 2 shows the modulated input (left) and the smoothed output laser beams (right) with two-dimensional and one-dimensional (the intensity profiles of central lines in the two two-dimensional spatial profiles) spatial profiles. As for the input laser beam, the spatial intensity modulation was larger than 2.4, and the modulation spatial frequency was 6 mm$^{-1}$. The gold-coated grating was 1400 g/mm with a size of about 1600×1000 mm$^2$, and the incident angle was 57°. The distance between the two parallel gratings was set to be approximately 2.3 m, and the induced spectral dispersion corresponded to a 4 ns chirped pulse. A spatial dispersion width as large as 520 mm was induced to the output laser beam, causing the output laser beam to have an excellent smoothed profile with a spatial intensity modulation of less than 1.1. Those less than 1.1 smoothed spatial intensity modulation results were always repeated with different input spatial modulations and modulation spatial frequencies. Except for the smoothing of spatial intensity modulation, the hot spots in the far-field induced by the wavefront aberrations at middle/high spatial frequency during free propagation can also be smoothed effectively [13].

**2.3 Spectral cutting effect of the SSGP main compressor**

In comparison to the symmetric four-grating compressor, the induced spatial dispersion width on the second grating (G2) by the SSGP compressor was twice wider, implying a larger area close to the central region in the laser beam suffers spectral cutting on G2 because of the limited grating size. Figure 3(a) clearly shows that the top-hat region on G2 with full spectral bandwidth was larger for the four-grating compressor (red solid line) than for the SSGP compressor (blue solid line). The spectral-cutting-induced energy loss was higher for the SSGP compressor. For the laser beam on the two edges, the spectral cutting on G2 was the same half spectral cutting for both the four-grating compressor and SSGP compressor. As for a 10s-100s PW laser, the laser beam can be more than 500 mm, and the central top-hat region of the laser beam with full spectral bandwidth was still large enough. With the same laser parameters as above, 860×860 mm$^2$ laser beam input, the spectral-cutting-induced energy loss was approximately 7.8% and 1.8% for the SSGP and four-grating compressors, respectively, as shown in Fig. 3(a). Here, the area ratio in Fig. 3(a) between the SSGP compressor (blue solid line) and the four-grating compressor (red solid line) was 94%. The area with full spectral bandwidth for the SSGP compressor shown in Fig. 3(a) was about 42%. The ideal two-dimensional beam profiles on G2 for the SSGP compressor and four-grating compressor are shown in Fig.3 (c) and 3(d), respectively. A For the 860×860 mm$^2$ input laser beam, there was light block on the edges of G1 and G2 in the four-grating compressor due to the relatively short grating pair distance.

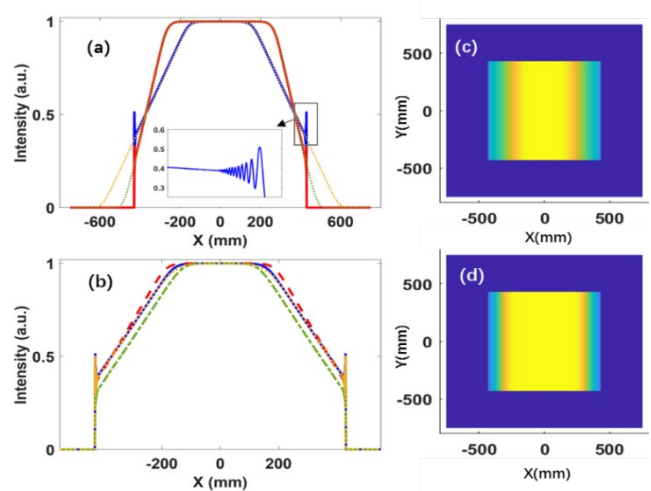

Fig.3. One-dimensional (a) (blue line and red line) or two-dimensional (c)(d) spatial profiles on G2 for SSGP compressor and four-grating compressor, respectively. The inset in (a) is the amplification of the edge cutting induced diffraction. (b) One-dimensional spatial profiles on G2 of SSGP compressor at different input conditions: original line (blue solid line), chirped-pulse is 3.5 ns (red dashed line), the 6$^{th}$

order super-Gaussian input beam (yellow dotted line), and spectral cutting at 40% (green dash-dot line), respectively.

According to the simulation, the inset in Fig. 3(a) showed the amplification of the edge cutting-induced diffraction, which did not induce serious spatial intensity modulation to the laser beam [13, 18]. Figure 3(b) shows the one-dimensional spatial profiles on G2 of the SSGP main compressor under different input conditions. The blue solid line shows the original line of the $10^{th}$ order super-Gaussian input beam, 4 ns chirped pulse duration, and spectral cutting at 50%. The red dashed line shows a 3.5 ns chirped pulse with shorter grating pair distance, indicating that a shorter chirped- pulse induces a smaller loss and wider central region. As for the $6^{th}$ order super-Gaussian input beam, shown by the yellow dotted line, the results were almost the same. The spectral cutting at 40%, shown with the green dash-dot line, implies a reduction in the energy loss and the top-hat length.

Although the spectral cutting induced energy loss was increased for the SSGP main compressor, because there were only two gratings used in the SSGP main compressor, the diffraction loss was greatly reduced compared to that of the four-grating main compressor. Assuming a diffraction efficiency of 90% for every grating, the total diffraction loss for an SSGP main compressor was 19.0%, while it was 34.4% for a four-grating main compressor. Assuming a spectral cutting loss of approximately 7.8% and 1.8% for the SSGP compressor and four-grating compressor, the total compression efficiencies for the SSGP and four-grating compressors were 74.7% and 64.4%, respectively. In other words, the compression efficiency was improved by 1.16 times (74.7/64.4) using the SSGP main compressor compared to that of the four-grating compressor. As a result, to achieve the same 10s-100s PW laser, the required maximum amplified chirped-pulse energy after the main amplifier can be reduced by a factor of approximately 1.16 times.

## 3. MS-MPC with SSGP main compressor

### 3.1 Description of the SSGP-main-compressor-based MS-MPC

We proposed a feasible design based on the MS-MPC with the SSGP main compressor to achieve compressed 10s-100s PW high peak power laser safely with a single laser beam, w. The optical diagrammatic sketch of the entire SSGP-main-compressor-based MS-MPC is shown in Fig.4.

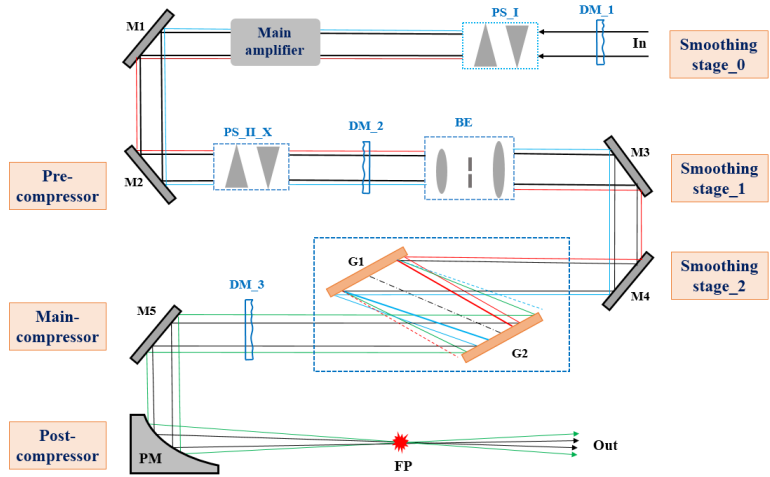

Fig. 4. Optical diagrammatic sketch of the SSGP-main-compressor-based MS-MPC. PS_I, PS_II_X: prism pair system. DM_1, DM_2, DM_3: reflective deformable mirrors with different size. BE: combined beam expander and relay imaging systems. G1, G2: diffraction gratings. M1-M5: plane reflective mirrors. PM: parabolic reflective mirror. FP: focal point.

The beam-smoothing process included three stages. Stage_0 induced a spatial dispersion width of about 2 to 3 mm in the horizontal direction using a prism pair PS_I to smoothen the incident laser beam before the main amplifier. All the potential hot spots induced damage risk to the main amplification crystal because spatial intensity modulation and wavefront aberration at high spatial frequencies were insulated. As for a super-Gaussian laser beam of several hundred millimeters, the effect of the induced spatial dispersion width of several millimeters on the amplification process can be neglected. In smoothing stage_1, a prism pair PS_II_X induces a suitable spatial dispersion width of less than 10 mm in the horizontal direction to smoothen the laser beam after the main amplifier, this smoothing setup protected the deformable mirror DM_2. Together with the following spatial-filter-based beam expander, BE, most of the wavefront aberration with high spatial frequency was filtered out, while the rest of the wavefront aberration with high/middle spatial frequency was smoothed. Smoothing stage_1 mainly protected the large optics, especially the first grating G1 and DM_2, which is also the pre-compressor stage. In the smoothing stage_2, which is also the main compressor, the grating pair composed of G1 and G2 induced a spatial dispersion width of hundreds of millimeters to smooth both the laser beam on G2 and the output beam. Then, it protected the last grating, G2, the weakest grating because of the shortest pulse duration in the central region with full spectral bandwidth. This extremely smoothed laser on spatial intensity and wavefront aberration also protected the final deformable mirror, DM_3, directly after G2. Finally, a simple parabolic reflective mirror was used as the post-compressor to compensate for the previously induced spatial dispersion at the focal point by using the spatiotemporal focusing effect [11, 19, 20].

**3.2 Spatiotemporal aberration compensation**

According to several theoretical and experimental studies, the spatiotemporal aberration induced by the large gratings in the four-grating main compressor, mainly the second grating, G2, and the third grating, G3, will affect both the pulse duration and the focal diameter and then decrease the focal intensity by almost 2 times [14-16]. In a four-grating compressor, the induced spatiotemporal aberration cannot be compensated using a deformable mirror before or after the compressor. To compensate this spatiotemporal distortion, a transmitted compensation plate was proposed, located between G2 and G3 [11, 21], to partly compensate for the static spatiotemporal aberration induced by G2 and G3. However, the compensation plate cannot compensate for the dynamic aberrations such as the heat-induced spatiotemporal distortion in the compressor [15]. Another indirect compensation method is using a deformable mirror based pre-compressor after the stretcher with a small beam size [22], which is not a direct method.

In the MS-MPC with the SSGP main compressor, three deformable mirrors were used to compensate for the wavefront aberration and the spatiotemporal distortion, as shown in Fig. 4. The first deformable mirror, DM_1, is used to ensure that the wavefront before PS_II_X was flat and well compensated. Here, the reflective light from the first prism was used as the wavefront monitoring light after beam reduction and then fed back to DM_1, as shown in Fig. 5(a).

After PS_II_X, the second deformable mirror, DM_2, was used to compensate for the induced wavefront aberration by the optics from PS_II_X to G1 (including). Here, the reflective light from G1 was used for wavefront monitoring after beam reduction and then fed back to DM_2, as shown in Fig. 5(b). As the SSGP was located in a vacuum chamber, window W2 was used to guide the reflective light from G1 out of the vacuum chamber. To remove the wavefront aberration in the measurement optical path, a removable reference flat mirror (RM) with a perfect flat wavefront was located before W2 in the chamber to calibrate the total wavefront aberration from W2 to Shack-Hartmann in Fig.5 (b), including the PM. This wavefront aberration can then be directly measured by the Shack-Hartmann with a light source illuminated at the location of the charge-coupled device (CCD). A standard laser interferometer, such as ZYGO, can also be used to precisely measure wavefront aberration. With a known wavefront aberration of the entire measurement optical path shown in Fig. 5 (b), the laser beam reflected output from G1 was expected to have well compensated flat wavefront, a similar wavefront aberration calibration procedure shown in Fig. 5(a). Note that the diffraction wavefront of the gratings included two parts: the reflective wavefront aberration and the static wavefront distortion induced by the error or imperfect grating lines, where the static wavefront distortion can be well measured offline. In this way, the diffractive laser beam output from G1 was also expected to have a compensated flat wavefront. Moreover, the monitoring lights for both DM_1 and DM_2 were not on the main beam line, indicating that both deformable mirrors can compensate the wavefront in real time.

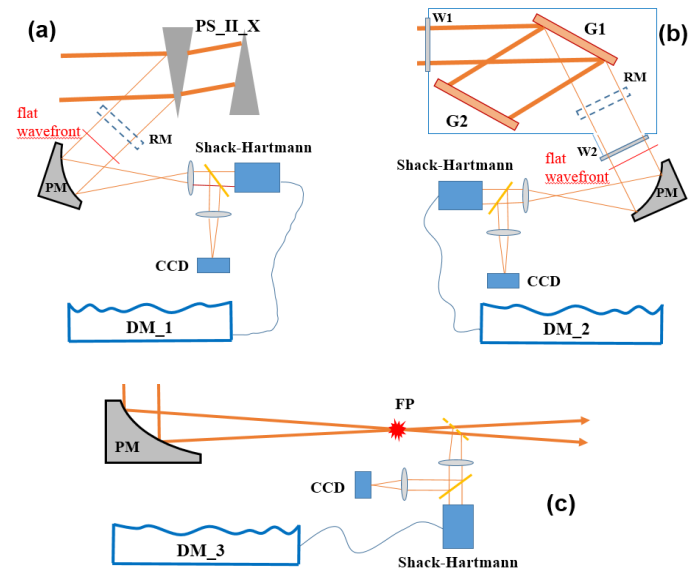

Fig. 5. Optical diagrammatic sketch of the wavefront distortion measuring and controlling for (a) DM_1, (b) DM_2, and (c) DM_3, respectively. DM_1, DM_2, DM_3: the reflective deformable mirrors corresponding to Fig. 4. PM: parabolic reflective mirror. W1, W2: windows. RM: reference flat mirror.

The laser beam after G1 with angular dispersion was collimated by using a G2. As the output beam from G2 was a collimated light with spatial dispersion, all the wavefront aberrations induced by G2 and the optics after G2 can be compensated effectively by using the third reflective deformable mirror DM_3, located directly after the MS-MPC. Wavefront monitoring for DM_3 was located near the final focal point to ensure the best focal intensity, as shown in Fig. 5(c). In this way, all the wavefront aberration and spatiotemporal distortion with low spatial frequency can be completely compensated in the SSGP based MS-MPC. As a result, it was expected to obtain a much higher focal intensity compared to that of the four-grating-based main compressor. The laser beam used for wavefront monitoring in Fig. 5(a) and (b) can simultaneously monitor the laser spectrum, pointing stability, and energy stability simultaneously, thus improving the operating efficiency of the laser facility.

**3.3 Spatiotemporal focusing properties**

Although the spatiotemporal aberration can be compensated effectively in the SSGP compressor, the compressor will induce a spatial dispersion width of hundreds of millimeters to the output laser beam, affecting the spatiotemporal properties at the focal point [11, 19, 20]. With the same input laser and optics parameters as in the above sections, the simulated output laser beam spatiotemporal properties at the focal point are shown in Fig. 6. The 14.3 fs transform limited pulse duration with 200 nm spectral bandwidth was broadened by about 0.7 fs and 1.6 fs to about 15.0 fs and 15.9 fs for a four-grating compressor and SSGP compressor, respectively. In the spatial domain, the focal diameters were almost the same for the three different conditions shown in Fig.6.

Furthermore, the influences of the beam diameter, spectral bandwidth, and ratio of the spatial dispersion region to the pulse duration on the focal point were also calculated and analyzed, as shown in Fig.7. According to Fig. 7(a), the focused pulse duration was broadened as the induced spatial dispersion width increased, while it decreased as the beam diameter increased. Figure 7(b) shows that the focused pulse duration was mainly related to the ratio of the spatial dispersion width to the laser beam. Figures 7(c) and (d) show that the spatial dispersion induced focused pulse duration broadening was also slightly increased as the spectral bandwidth increased, indicating that this SSGP was more suitable for high peak power lasers with narrow spectral bandwidths.

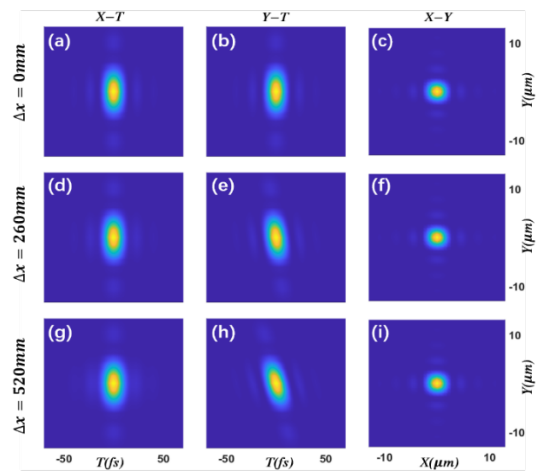

Fig. 6. Spatiotemporal properties at the focal plane with and without spatial dispersion on the laser beam. Spatiotemporal properties and the integrated spatial profiles at the focal plane in the case of inducing (a-c) 0 mm, (d-f) 260 mm, and (g-i) 520 mm spatial dispersion width, respectively.

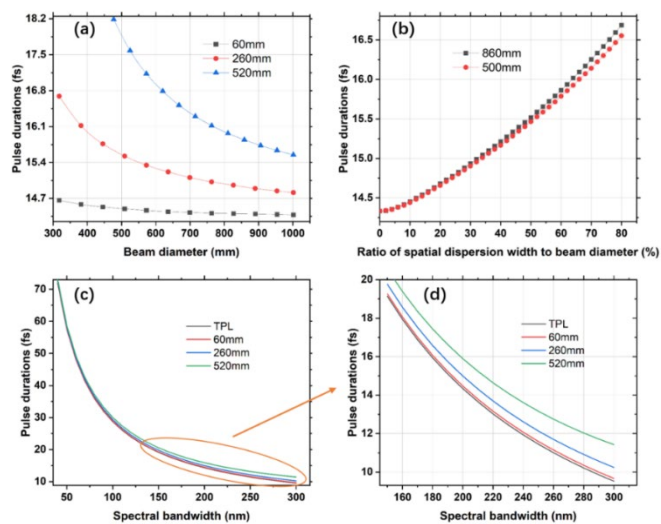

Fig. 7. Curves of the pulse duration on the focal point related to (a) the beam diameter, (b) the ratio of spatial dispersion region, and (c)(d) the spectral bandwidth, at different induced spatial dispersion widths (TPL=0, 60, 260, 520 mm) or beam size (500mm, 860mm), respectively.

## 4. Proof-of-principle experiment and 50-100 PW design

A two-grating compressor was set up after a kHz Ti:sapphire amplifier (Coherent, Legend) to experimentally prove the feasibility of the SSGP main compressor, a. The full spectral bandwidth ranged from 760 nm to 830 nm with a Gaussian spectral profile. The two compression gratings were home-made 1480 g/mm gold-coated grating with a size of 150 mm × 200 mm, with an incident angle of 56º. The distance of between the two gratings was separated by approximately 600 mm to fully compensate the spectral dispersion, inducing a spatial dispersion width of about 42 mm to the output laser beam, as shown in Fig. 8(a). For a 50-mm-diameter input laser beam, the central part with full spectral bandwidth of the output laser beam was 8 mm, indicating that the ratio of the spatial dispersion width to the laser beam (Fig. 7(b)) was approximately 90%.

There is no doubt that the output laser beam after the SSGP compressor was smoothed effectively, as shown in Fig. 8(b), when there was clear induced spatial intensity modulation on the laser beam before G1 (Fig. 8(a)) or between G1 and G2. After the SSGP compressor, the temporal profile at the focal point was measured by using the SHG-FROG method with retrieval error of less than 0.005, as shown in Fig. 8(c). As the laser spectral bandwidth was relatively narrow, even though the ratio of the spatial dispersion width to the laser beam was about 90%, the pulse duration after the SSGP compressor was almost the same as the pulse duration obtained by using a double-pass SSGP compressor with no spatial dispersion. Figure 8(d) shows the focal spot of the output laser beam focused by using a large lens with a diameter of 220 mm and focal length of about 4 m, which also shows a good focus. The temporal and spatial profiles at the focal point in the experiment clearly demonstrate the feasibility of the SSGP main compressor. The SSGP compressor has been successfully used in two-photon microscopy and ultrafast micromachining to obtain a better optical section effect or higher axial resolution [19, 20]. In the PETAL PW facility, the entire compressor was composed of a symmetric four-grating compressor and a final similar SSGP compressor, which proving that the proposed SSGP-based MS-MPC is feasible for a PW laser system [23].

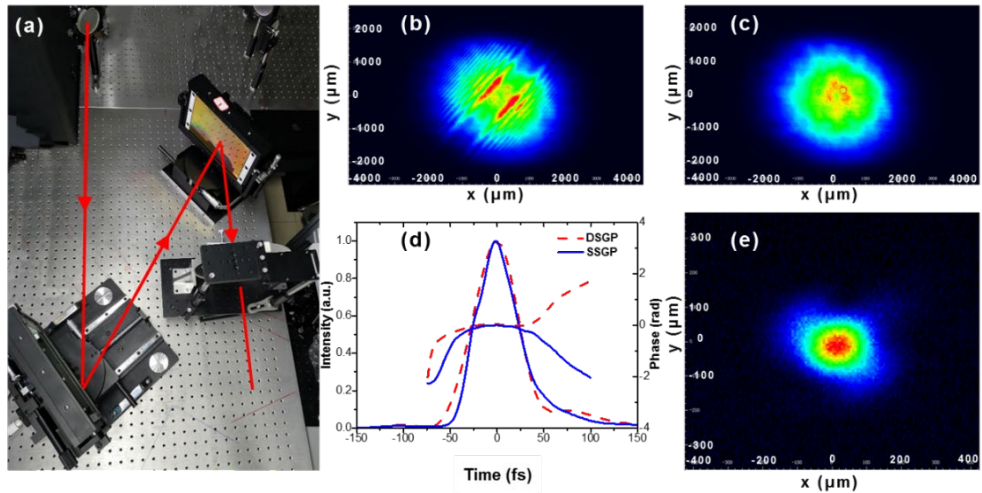

Fig. 8. (a) Experimental setup of a SSGP main compressor; (b) spatial intensity modulations on the laser beam before G1; (c) output laser beam after the SSGP compressor; (d) temporal profiles at the focal point for single-pass and double-pass single-grating pair; and (e) focal spot of the output laser beam.

The optical diagrammatic sketch, shown in Fig.4, can be used to design a 50-100 PW laser output by using the SSGP-based MS-MPC. For gold-coated gratings, the damage threshold is about 0.2 J/cm$^2$ for the femtosecond laser on the last grating, which limits the maximum compressed output pulse energy or peak power [11, 21]. As the spatial intensity modulation of the laser beam can be smoothed to less than 1.1 times owing to the induced hundreds of millimeters spatial dispersion width to the output laser beam, $\sqrt{2}$ times spatial intensity modulation was used instead of typical 2.0 times. This setting can keep the last grating safe, corresponding to the maximum energy density on the last grating of 0.141 J/cm$^2$. Here, the energy density distribution on G2 was used as the calculation reference. For a commercially available grating with an effective area of approximately 1400×660 mm$^2$, the maximum compressed output pulse energy should be about 140×65×0.141×80%=1026 J, when considering the approximately 1-80%=20% non-full-filled area on both sides of the blue line, similar to Fig. 3(a). After spatiotemporal focusing with about 16.6 fs at the focal point according to Fig. 7(a) with about 650 mm beam size, the corresponding focal peak power was about 1026J/ 16.6 fs=61.8 PW. For a gold-coated grating with an effective area of about 1600×1000 mm$^2$, the maximum compressed output pulse energy should be about 160×95×0.141×80% =1715 J, where 80% is also the ratio of the blue line area, as shown in Fig. 3(a), to a rectangular area of about 900×1 mm$^2$. According to Fig. 7(a), spatiotemporal focusing with about 15.7 fs at the focal point resulted in about 1715 J/ 15.7 fs=109.2 PW.

## 5. Conclusion

In conclusion, an improved MS-MPC optical design based on an SSGP main compressor was studied systematically. In comparison to the previous MS-MPC based on AFGC, the SSGP main compressor-based MS-MPC shows lots of advantages, such as saving for two expensive gratings, simplifying the setup, simplifying the vacuum chamber of the main compressor, improving the beam smoothing effect of the output laser beam, and increasing the total compression efficiency. Furthermore, the spatiotemporal aberration induced by the gratings can be compensated effectively by using several deformable mirrors, which can result in a higher focal intensity when considering the spatiotemporal effect in MS-MPC based on AFGC. This property can also reduce the wavefront aberration requirement of large gratings, which are difficult or expensive to be controlled effectively during the manufacturing and installation processes. As the two-grating distance is long enough in the SSGP main compressor, a shorter 3.0 ns or 3.5 ns chirped-pulse instead of a 4 ns chirped-pulse, or typical 1480 g/mm gratings are feasible in this design, thus extending the choice range of many parameters to optimize the compressor further. All the above advantages make it a promising method for PW laser compressors with relatively large beam sizes. In the near future, by using this SSGP-based MS-MPC with a single laser beam, approximately 60 PW or 100 PW laser can be obtained by using only two 1400×660 mm$^2$ or 1600×1000 mm$^2$ gold-coated gratings, respectively.

**Funding.** This work was supported by the National Natural Science Foundation of China (NSFC) (61527821, 61905257, U1930115) and the Shanghai Municipal Science and Technology Major Project (2017SHZDZX02).

**Acknowledgments.** The authors would like to thank Dr. Yunxia Jin for supplying gratings for the experiment, and Dr. Chenqiang Zhao, and Dr. Lianghong Yu for the discussion.

**Disclosures.** The authors declare no conflicts of interest.

**References**
1. T. H. Maiman, "Stimulated optical radiation in ruby," Nature **187**(4736), 493–494 (1960).
2. D. Strickland and G. Mourou, "Compression of amplified chirped optical pulses," Optics Communications **55**, 219-221 (1985).
3. A. Dubietis, G. Jonusauskas, and A. Piskarskas, "Powerful femtosecond pulse generation by chirped and stretched pulse parametric amplification in BBO crystal," Optics Communications **88**, 437-440 (1992).
4. Z. Gan, L. Yu, C. Wang, Y. Liu, Y. Xu, W. Li, S. Li, L. Yu, X. Wang, X. Liu, J. Chen, Y. Peng, L. Xu, B. Yao, X. Zhang, L. Chen, Y. Tang, X. Wang, D. Yin, X. Liang, Y. Leng, R. Li, and Z. Xu, "The Shanghai Superintense Ultrafast Laser Facility (SULF) Project," in Progress in Ultrafast Intense Laser Science XVI, K. Yamanouchi, K. Midorikawa, and L. Roso, Eds. (2021), pp. 199-217.
5. F. Lureau, G. Matras, O. Chalus, C. Derycke, T. Morbieu, C. Radier, O. Casagrande, S. Laux, S. Ricaud, G. Rey, A. Pellegrina, C. Richard, L. Boudjemaa, C. Simon-Boisson, A. Baleanu, R. Banici, A. Gradinariu, C. Caldararu, B. De Boisdeffre, P. Ghenuche, A. Naziru, G. Kolliopoulos, L. Neagu, R. Dabu, I. Dancus, and D. Ursescu, "High-energy hybrid femtosecond laser system demonstrating 2 x 10 PW capability," High Power Laser Science and Engineering **8**, e43 (2020).
6. Z. X. Zhang, F. X. Wu, J. B. Hu, X. J. Yang, J. Y. Gui, P. H. Ji, Y. Q. Liu, C. Wang, Y.Q.Liu, X. M. Lu, Y. Xu, Y. X. Leng, R. X. Li, and Z. Z. Xu, "The 1PW/0.1Hz laser beamline in SULF facility," High Power Laser Science and Engineering, 8, E4 (2020).
7. C. N. Danson, M. White, J. R. M. Barr, T. Bett, P. Blyth, D. Bowley, C. Brenner, R. J. Collins, N. Croxford, A. E. B. Dangor, L. Devereux, P. E. Dyer, A. Dymoke-Bradshaw, C. B. Edwards, P. Ewart, A. I. Ferguson, J. M. Girkin, D. R. Hall, D. C. Hanna, W. Harris, D. I. Hillier, C. J. Hooker, S. M. Hooker, N. Hopps, J. Hull, D. Hunt, D. A. Jaroszynski, M. Kempenaars, H. Kessler, P. L. Knight, S. Knight, A. Knowles, C. L. S. Lewis, K. S. Lipton, A. Littlechild, J. Littlechild, P. Maggs, G. P. A. Malcolm, S. P. D. Mangles, W. Martin, P. McKenna, R. O. Moore, C. Morrison, Z. Najmudin, D. Neely, G. H. C. New, M. J. Norman, T. Paine, A. W. Parker, R. R. Penman, G. J. Pert, C. Pietraszewski, A. Randewich, N. H. Rizvi, N. Seddon, Z. M. Sheng, D. Slater, R. A. Smith, C. Spindloe, R. Taylor, G. Thomas, J. W. G. Tisch, J. S. Wark, C. Webb, S. M. Wiggins, D. Willford, and T. Winstone, "A history of high-power laser research and development in the United Kingdom," High Power Laser Science and Engineering 9(2021)
8. C. N. Danson, C. Haefner, J. Bromage, T. Butcher, J. Chanteloup, E. A. Chowdhury, A. Galvanauskas, L. A. Gizzi, J. Hein, and D. I. Hillier, "Petawatt and exawatt class lasers worldwide," High Power Laser Science and Engineering **7**, e54 (2019).
9. A. Di Piazza, C. Muller, K. Z. Hatsagortsyan, and C. H. Keitel, "Extremely high-intensity laser interactions with fundamental quantum systems," Reviews of Modern Physics **84**, 1177-1228 (2012).
10. D. Wang and Y. X. Leng, "Simulating a four-channel coherent beam combination system for femtosecond multi-petawatt lasers," Optics Express **27**, 36137-36153 (2019).
11. J. Liu, X. Shen, S. M. Du, and R. X. Li, "Multistep pulse compressor for 10s to 100s PW lasers," Optics Express **29**, 17140-17158 (2021).

12. X. Shen, S. M. Du, J. Liu, and R. X. Li, "Asymmetric four-grating compressor for ultrafast high power lasers," arXiv: 2105.04863 (2021).
13. S. M. Du, X. Shen, W. H. Liang, P. Wang, J. Liu, and R. X. Li, "Multistage smoothing based multistep pulse compressor for ultrahigh peak power lasers," arXiv:2201.03917 (2022).
14. Z. Li and N. Miyanaga, "Simulating ultra-intense femtosecond lasers in the 3-dimensional space-time domain," Optics Express **26**, 8453-8469 (2018).
15. V. Leroux, T. Eichner, and A. R. Maier, "Description of spatio-temporal couplings from heat-induced compressor grating deformation," Optics Express **28**, 8257-8265 (2020).
16. G. Pariente, V. Gallet, A. Borot, O. Gobert and F. Quéré, "Space–time characterization of ultra-intense femtosecond laser beams," Nature Photonics **10**(8), 547-553 (2016).
17. E. Treacy, "Optical pulse compression with diffraction gratings," IEEE Journal Quantum Electronics **5**(9), 454-458 (1969).
18. Y. Ohtsuka and M. C. Yin, "Fresnel diffraction by a semitransparent straight edge object with acoustically coherence-controllable illumination," Applied Optics 23, 300-306 (1984).
19. G. Zhu, J. V. Howe, M. Durst, W. Zipfel, and C. Xu, "Simultaneous spatial and temporal focusing of femtosecond pulses," Optics Express **13**, 2153-2159 (2005).
20. F. He, B. Zeng, W. Chu, J. Ni, K. Sugioka, Y. Cheng, and C. Durfee, "Characterization and control of peak intensity distribution at the focus of a spatiotemporally focused femtosecond laser beam," Optics Express **22**, 9734–9748 (2014).
21. J. Liu, X. Shen, Z. Si, C. Wang, C. Q. Zhao, X. Y. Liang, Y. X. Leng and R. X. Li, "In-house beam-splitting pulse compressor for high-energy petawatt lasers," Optics Express **28**(15), 22978-22991 (2020).
22. Z. Li and J. Kawanaka, "Complex spatiotemporal coupling distortion pre-compensation with double-compressors for an ultra-intense femtosecond laser," Optics Express **27**, 25172-25186 (2019).
23. N. Blanchot, G. Béhar, J.C. Chapuis, C. Chappuis, S. Chardavoine, J.F. Charrier, H. Coïc, C. Damiens-Dupont, J. Duthu, P. Garcia, J. P. Goossens, F. Granet, C. Grosset-Grange, P. Guerin, B. Hebrard, L. Hilsz, L. Lamaignere, T. Lacombe, E. Lavastre, T. Longhi, J. Luce, F. Macias, M. Mangeant, E. Mazataud, B. Minou, T. Morgaint, S. Noailles, J. Neauport, P. Patelli, E. Perrot-Minnot, C. Present, B. Remy, C. Rouyer, N. Santacreu, M. Sozet, D. Valla, and F. Laniesse, "1.15 PW–850 J compressed beam demonstration using the PETAL facility," Optics Express **25**, 16957-16970 (2017).